
\hsize=6.0in
\hoffset=0.25in
\tolerance=1000\hfuzz=2pt


\def\monthintext{\ifcase\month\or January\or February\or
   March\or April\or May\or June\or July\or August\or
   September\or October\or November\or December\fi}




\def\date#1{\leftline{#1}}
\def\preprint#1#2#3{\baselineskip=18pt\pageno=0
   \begingroup
   \vsize=8.0in\voffset=0.5in\nopagenumbers\parindent=0pt\baselineskip=14.4pt
   \rightline{#1}\rightline{#2}\rightline{#3}}
\def\title#1#2{
   \vskip 0.7in plus 0.2in
   \centerline{\titlefont #1}
   \vskip 0.2in
   \centerline{\titlefont #2}
   \vskip 0.8in plus 0.1in}

\def\author#1#2#3{\centerline{{\bf #1}\ulabelfoot{#2}{#3}}\smallskip}
\def\address#1{\centerline{#1}}
\def\abstract{\bigskip\bigskip\bigskip\medskip
	\centerline{\bf Abstract}
	\smallskip}
\def\finishtitlepage{\vskip 0.8in plus 0.3in
   \supereject\endgroup
   \baselineskip=18pt}

\def\nolabels{\def\eqnlabel##1{}\def\eqlabel##1{}\def\figlabel##1{}%
	\def\reflabel##1{}}
\def\writelabels{\def\eqnlabel##1{%
	{\escapechar=` \hfill\rlap{\hskip.11in\string##1}}}%
	\def\eqlabel##1{{\escapechar=` \rlap{\hskip.11in\string##1}}}%
	\def\figlabel##1{\noexpand\llap{\string\string\string##1\hskip.66in}}%
	\def\reflabel##1{\noexpand\llap{\string\string\string##1\hskip.37in}}}
\nolabels


\global\newcount\secno \global\secno=0
\global\newcount\meqno \global\meqno=1

\def\newsec#1{\global\advance\secno by1
\xdef\secsym{\the\secno.}\global\meqno=1
	\bigbreak\medskip
	\noindent{\bf\the\secno. #1}\par\nobreak\medskip\nobreak\noindent}
\xdef\secsym{}

\def\appendix#1#2{\global\meqno=1\xdef\secsym{\hbox{#1.}}\bigbreak\bigskip
\noindent{\bf Appendix #1. #2}\par\nobreak\medskip\nobreak}


\def\eqnn#1{\xdef #1{(\secsym\the\meqno)}%
	\global\advance\meqno by1\eqnlabel#1}
\def\eqna#1{\xdef #1##1{\hbox{$(\secsym\the\meqno##1)$}}%
	\global\advance\meqno by1\eqnlabel{#1$\{\}$}}
\def\eqn#1#2{\xdef #1{(\secsym\the\meqno)}\global\advance\meqno by1%
	$$#2\eqno#1\eqlabel#1$$}


\def\ulabelfoot#1#2{{\baselineskip=14.4pt plus 0.3pt\footnote{#1}{#2}}}
\global\newcount\ftno \global\ftno=1
\def\foot#1{{\baselineskip=14.4pt plus 0.3pt\footnote{$^{\the\ftno}$}{#1}}%
	\global\advance\ftno by1}


\global\newcount\refno \global\refno=1
\newwrite\rfile

\def\ref{[\the\refno]\nref}
\def\nref#1{\xdef#1{[\the\refno]}\ifnum\refno=1\immediate
	\openout\rfile=refs.tmp\fi\global\advance\refno by1\chardef\wfile=\rfile
	\immediate\write\rfile{\noexpand\item{#1\ }\reflabel{#1}\pctsign}\findarg}
\def\findarg#1#{\begingroup\obeylines\newlinechar=`\^^M\passarg}
	{\obeylines\gdef\passarg#1{\writeline\relax #1^^M\hbox{}^^M}%
	\gdef\writeline#1^^M{\expandafter\toks0\expandafter{\striprelax #1}%
	\edef\next{\the\toks0}\ifx\next\null\let\next=\endgroup\else\ifx\next\empty%
	\else\immediate\write\wfile{\the\toks0}\fi\let\next=\writeline\fi\next\relax}}
	{\catcode`\%=12\xdef\pctsign{
\def\striprelax#1{}

\def\semi{;\hfil\break}
\def\addref#1{\immediate\write\rfile{\noexpand\item{}#1}} 

\def\listrefs{\vfill\eject\immediate\closeout\rfile
   \centerline{{\bf References}}\bigskip{\frenchspacing%
   \catcode`\@=11\escapechar=` %
   \input refs.tmp\vfill\eject}\nonfrenchspacing}

\def\startrefs#1{\immediate\openout\rfile=refs.tmp\refno=#1}


\global\newcount\figno \global\figno=1
\newwrite\ffile
\def\fig{\the\figno\nfig}
\def\nfig#1{\xdef#1{\the\figno}\ifnum\figno=1\immediate
	\openout\ffile=figs.tmp\fi\global\advance\figno by1\chardef\wfile=\ffile
	\immediate\write\ffile{\medskip\noexpand\item{Fig.\ #1:\ }%
	\figlabel{#1}\pctsign}\findarg}

\def\listfigs{\vfill\eject\immediate\closeout\ffile{\parindent48pt
	\baselineskip16.8pt\centerline{{\bf Figure Captions}}\medskip
	\escapechar=` \input figs.tmp\vfill\eject}}



\font\titlerm=cmr10 scaled \magstep3
\font\titlerms=cmr10 scaled \magstep1

\font\titlei=cmmi10 scaled \magstep3  
\font\titleis=cmmi10	scaled \magstep1

\font\titlesy=cmsy10 scaled \magstep3 	
\font\titlesys=cmsy10 scaled \magstep1

\font\titleit=cmti10 scaled \magstep3	

\skewchar\titlei='177 \skewchar\titleis='177 
\skewchar\titlesy='60 \skewchar\titlesys='60 

\def\titlefont{\def\rm{\fam0\titlerm}
   \textfont0=\titlerm \scriptfont0=\titlerms 
   \textfont1=\titlei  \scriptfont1=\titleis  
   \textfont2=\titlesy \scriptfont2=\titlesys 
   \textfont\itfam=\titleit \def\it{\fam\itfam\titleit} \rm}



\font\tenrm=cmr10 scaled \magstep1
\font\sevenrm=cmr7 scaled \magstep1
\font\fiverm=cmr5 scaled \magstep1

\font\tenbf=cmbx10 scaled \magstep1
\font\sevenbf=cmbx7 scaled \magstep1
\font\fivebf=cmbx5 scaled \magstep1

\font\teni=cmmi10 scaled \magstep1
\font\seveni=cmmi7 scaled \magstep1
\font\fivei=cmmi5 scaled \magstep1

\font\tensy=cmsy10 scaled \magstep1
\font\sevensy=cmsy7 scaled \magstep1
\font\fivesy=cmsy5 scaled \magstep1

\font\tenex=cmex10 scaled \magstep1
\font\tentt=cmtt10 scaled \magstep1
\font\tenit=cmti10 scaled \magstep1
\font\tensl=cmsl10 scaled \magstep1

\def\twelvepoint{\def\rm{\fam0\tenrm}
	\textfont0=\tenrm \scriptfont0=\sevenrm \scriptscriptfont0=\fiverm
	\textfont1=\teni  \scriptfont1=\seveni  \scriptscriptfont1=\fivei
	\textfont2=\tensy \scriptfont2=\sevensy \scriptscriptfont2=\fivesy
	\textfont\itfam=\tenit \def\it{\fam\itfam\tenit}
	\textfont\ttfam=\tentt \def\tt{\fam\ttfam\tentt}
	\textfont\bffam=\tenbf \def\bf{\fam\bffam\tenbf}
	\textfont\slfam=\tensl \def\sl{\fam\slfam\tensl} \rm
   \hfuzz=1pt\vfuzz=1pt
   \setbox\strutbox=\hbox{\vrule height 10.2pt depth 4.2pt width 0pt}
   \parindent=24pt\parskip=0pt plus 1.2pt
   \topskip=12pt\maxdepth=4.8pt\jot=3.6pt
	\normalbaselineskip=14.4pt\normallineskip=1.2pt
   \normallineskiplimit=0pt\normalbaselines
	\abovedisplayskip=13pt plus 3.6pt minus 5.8pt
   \belowdisplayskip=13pt plus 3.6pt minus 5.8pt
   \abovedisplayshortskip=-1.4pt plus 3.6pt
   \belowdisplayshortskip=13pt plus 3.6pt minus 3.6pt
   \topskip=12pt \splittopskip=12pt
   \scriptspace=0.6pt\nulldelimiterspace=1.44pt\delimitershortfall=6pt
   \thinmuskip=3.6mu\medmuskip=3.6mu plus 1.2mu minus 1.2mu
   \thickmuskip=4mu plus 2mu minus 1mu
   \smallskipamount=3.6pt plus 1.2pt minus 1.2pt
   \medskipamount=7.2pt plus 2.4pt minus 2.4pt
   \bigskipamount=14.4pt plus 4.8pt minus 4.8pt}

\twelvepoint


\def\noblackbox{\overfullrule=0pt}
\def\inv{^{\raise.18ex\hbox{${\scriptscriptstyle -}$}\kern-.06em 1}}
\def\dup{^{\vphantom{1}}}
\def\Dsl{\,\raise.18ex\hbox{/}\mkern-16.2mu D} 
\def\dsl{\raise.18ex\hbox{/}\kern-.68em\partial}
\def\slash#1{\raise.18ex\hbox{/}\kern-.68em #1}
\def\boxeqn#1{\vcenter{\vbox{\hrule\hbox{\vrule\kern3.6pt\vbox{\kern3.6pt
	\hbox{${\displaystyle #1}$}\kern3.6pt}\kern3.6pt\vrule}\hrule}}}
\def\mbox#1#2{\vcenter{\hrule \hbox{\vrule height#2.4in
	\kern#1.2in \vrule} \hrule}}  
\def\half{{\textstyle{1\over2}}}
\def\ha{{1\over2}}
\def\e#1{{\rm e}^{\textstyle#1}}
\def\C#1{\hbox{{$\cal #1$}}}    
\def\O{\hbox{{$\cal O$}}}	
\def\({\left(} \def\){\right)}  
\def\[{\left[} \def\]{\right]}  
\def\l|{\left|} \def\r|{\right|}
\def\vev#1{\langle #1 \rangle}
\def\<{\langle}
\def\>{\rangle}
\def\psibar{\overline\psi}
\def\lform{\hbox{$\sqcup$}\llap{\hbox{$\sqcap$}}}
\def\grad#1{\,\nabla\!_{{#1}}\,}
\def\gradgrad#1#2{\,\nabla\!_{{#1}}\nabla\!_{{#2}}\,}
\def\darr#1{\raise1.8ex\hbox{$\leftrightarrow$}\mkern-19.8mu #1}
\def\roughly#1{\raise.3ex\hbox{$#1$\kern-.75em\lower1ex\hbox{$\sim$}}}
\hyphenation{di-men-sion di-men-sion-al di-men-sion-al-ly
             di-men-sion-al-i-ty}

\noblackbox
\def\suc#1{\hbox{$SU(#1)_c$}}
\def\su#1{\hbox{$SU(#1)$}}
\def\sul{\hbox{$SU(2)_L$}}
\def\fred{\hbox{$SU(2)'$}}
\def\mzp{\hbox{$M_{Z^\prime}$}}
\def\sm{standard model} \def\Sm{Standard model}
\def\ss{\scriptscriptstyle}

\date{February 1992}
\preprint{McGill/92-06}{hep-ph@xxx/9203205}{}
\title{Extended Color Models with a Heavy Top Quark }
{}
\author{Oscar F. Hern\'andez}{$^{\dagger}$}
{internet: oscarh@physics.mcgill.ca \ ,\ \ \ \ \ decnet: 19191::oscarh}
\address{Department of Physics, ERP Building}
\address{ 3600 University St. }
\address{McGill University}
\address{Montr\'eal, Qu\'ebec, Canada H3A 2T8}
\abstract

We present a class of models in which the top quark, by mixing with new physics
at a higher energy scale, is naturally heavier than the other \sm\ particles.
We take this new physics to be extended color.  Our models contain
new particles with masses between 100 GeV and 1 TeV, some
of which may be just within the reach of the next generation of experiments.
In particular one of our models implies the existence of two right handed top
quarks.  These models demonstrate the existence of a \sm-like theory consistent
with experiment, and leading to new physics below the TeV scale, in which the
third generation is treated differently than the first two.

\finishtitlepage
\newsec{Introduction}  \noindent   Upon examining the fundamental particle
spectrum one is immediately impressed by how much heavier the top quark is
than all the other quarks.   
In the \sm\ this is achieved in an unnatural way.  One tunes most of the
Yukawa couplings to be very small and allows one of them to be of order one.
Here we propose a class of models in which  the much higher mass of the top
quark (with respect to the other fundamental  fermions) is a result of the
dynamics and symmetry breaking in the theory. The key feature of the models we
propose is that while the top quark gets its mass from electroweak symmetry
breaking, this quark also mixes with physics at a higher energy scale. We
propose the physics of this higher energy scale to be an extended color group.
In particular, we have in mind models like \suc4\ \ref\f{R. Foot, Phys. Rev. D
40, (1989) 3136.} and \suc 5\ color \ref\fh{R. Foot, O.F. Hern\'andez, Phys.
Rev. D 41 (1990) 2283.}.
\nref\gs{S.L. Glashow and U. Sarid, Phys. Lett. B 246 (1990) 188.}
\nref\r{T.~Rizzo, Univ.~of Wisc.-Madison preprint MAD/PH/626.}
\nref\fhri{R. Foot, O.F. Hern\'andez, T. Rizzo, Phys. Lett. B 246 (1990) 183.}
\nref\fhrii{R. Foot,  O.F.~Hern\'andez, T. Rizzo, Phys. Lett. B 261 (1991)
153.}
\nref\oh{O.F.~Hern\'andez, Phys. Rev. D 44 (1991) 1997.}
\nref\chsb{ E. Carlson, L. Hall, U. Sarid, J. Burton, Phys. Rev. D44 (1991)
1555.}
The \suc 5\ model was particularly successful in generating new phenomena at
the multi-GeV energy scale while agreeing with experiment to the same extent as
the \sm\  \gs-\chsb.

In the extended color model scenario, the constraint of reproducing the \sm\ at
energies below 100 GeV led to \suc 5\ as the only possiblility \fh,\ref\fhii{R.
Foot, O.F. Hern\'andez, Univ. of Wisc.-Madison preprint MAD/TH/90-10.}.  This
constraint meant that none of the \sm\ particles received a contribution to
their mass from the breaking of the extended color group.  We are no longer
restricted to the group \suc 5; however that is the group we will use to
describe our scenario. In section 2 we present the basic ideas behind two
different models that allow for a heavy top quark and discuss their common
features. Sections 3 and 4 concentrate on the specific features of each of the
two respective models.  Finally in section 5 we present our criticisms and
conclusions.

\newsec{The models} We would like to reformulate the \suc 5\ color model \fh\
in such a way that it leads to a heavy top quark and yet retains all its
previous successes.  Non-\sm\ alternatives to the color sector are especially
important avenues to explore.  For example, hadronic physics has nothing
comparable to the stunning theoretical agreement with precision LEP
measurements, whereas revising the leptonic sector would almost most certainly
contradict LEP results (or force us to place the new physics at a much higher
energy scale). Our models therefore always leave the leptonic sector unchanged.

One loophole which we exploit is that experiments have very little to say
about the right handed top quark, $t_R$, since it is an \sul\ singlet.
It is actually $t_R$ which we
will couple to the extended color physics coming from the higher energy scale.
This comes about because $t_R$ is in a higher dimensional
representation of \suc 5, whereas the other quarks are in the fundamental.
This is one of the major differences between these models and the original
\suc 5\ of reference~\fh.   We present two possiblities for the
\suc 5\ representation of $t_R$, the 10 and the $\overline{10}$.

Consider a theory with gauge group
\eqn\gaugegrp
{SU(5)_{\rm color}\otimes SU(2)_L \otimes U(1)_{y}'.
}
As stated above the leptonic sector is unchanged, and under \gaugegrp\ they
have the quantum numbers
\eqn\lepcontent{
f_L \sim (1, 2, -1),~~~ e_R \sim (1, 1, -2),
}
while the first two quark generations have the form
\eqn\quarkcontent{
Q_L \sim (5, 2, y_Q),~~~ u_R \sim (5, 1, y_Q+1),~~~d_R \sim (5, 1, y_Q-1) ,
}
just as in the original \suc 5\ color model \fh.
However the third quark generation, and in particular the \sul\ singlet top
quark, have different quantum numbers.  We will be studying two types of
models. The ``ten bar'' model has $t_R$ in the $\overline{10}$ of \suc5,
\eqn\thirdtenbar{
T_L \sim (5, 2, y_Q),~~~ t_R \sim (\overline{10}, 1, y_{t_R}),~~~%
b_R \sim (5, 1, y_Q-1 ) ,
}
and the ``ten'' model has $t_R$ in the 10 of \suc5,
\eqn\thirdtenbar{
T_L \sim (5, 2, y_Q),~~~ t_R \sim (10, 1, y_{t_R}),~~~%
b_R \sim (5, 1, y_Q-1 ) .
}

Just as in \fh\ we break \suc 5\ with the Higgs
\eqn\eqchi{
\chi\sim(10,1,2y_Q).
}
However we use a colored Higgs, $H$, to break the
electroweak symmetry, and give masses to the $W$ and $Z$ bosons and the top
quark.  Its quantum numbers are
\eqn\higgstenb{
H\sim (\overline{10},2,y_Q-y_{t_R}) ,
}
for $t_R\sim\overline{10}$ and
\eqn\higgsfive{
H\sim(\overline5,2,y_Q-y_{t_R}) ,
}
for $t_R\sim 10$.
The Yukawa Lagrangian is
\eqn\yuk
{
L_{\rm Yuk} = \lambda \overline T_L H t_R +
\lambda' \overline T_L \chi (T_L)^c +
\lambda_1 \overline Q_L \chi (Q_L)^c + \lambda_2 \overline u_R \chi (d_R)^c
+{\rm h.c.}\ .
}

Consider the \su 5\ tensor products
\eqn\tensor{\eqalign{
\overline{5}\times\overline{10} & = 10 + 40 \cr
\overline{5}\times 10           & = 5  + 45
,\cr}}
and the following \suc 5\ branching rules under $\fred
\times\suc3\times\tilde{U}(1)$.
\eqn\branching{\eqalign{
10= & (1,1)(6) + (1,\overline3)(-4) + (2,3)(1) \cr
40= & (2,1)(9) + (1,\overline3)(-4) + (2,3)(1) + (3,\overline3)(-4)+\cdots \cr
 5= & (2,1)(3) + (1,3)(-2) \cr
45= & (2,1)(3) + (1,3)(-2) + (3,3)(-2) + (1,\overline{3})(8)+ \cdots
. \cr}}
In order for the Higgs $H$ to give a mass to the top, the $\tilde{U}(1)$
charges must satisfy the constraint
$\tilde{y}_H-\tilde{y}_{T_L}+\tilde{y}_{t_R}=0$.
Eqs. \yuk, \tensor, \branching\ show that we have no
choice but to pick $H$ in the $\overline{10}$ for $t_R$ in the
$\overline{10}$.  If we take $t_R$ in the 10, we can pick $H$ to
be either in the $\overline5$ or $\overline{45}$ representation; however we
will only consider $H$ in the $\overline 5$.

The symmetry breaking pattern for the ten bar model is:
\eqn\symbrktb{
\eqalign{\suc5 \otimes & SU(2)_L \otimes U(1)_{y}'\cr
       &\downarrow \vev{\chi}\sim  w \cr
    SU(2)' \otimes SU(3)_c & \otimes SU(2)_L \otimes U(1)_y\cr
       &\downarrow  \vev{H}\sim u \cr
SU(2)' \otimes & SU(3)_c \otimes U(1)_{em}\cr}
}
That for the ten model is:
\eqn\symbrkt{
\eqalign{\suc5 \otimes & SU(2)_L \otimes U(1)_{y}'\cr
       &\downarrow \vev{\chi}\sim  w \cr
    SU(2)' \otimes SU(3)_c & \otimes SU(2)_L \otimes U(1)_y\cr
       &\downarrow  \vev{H}\sim u \cr
SU(3)_c & \otimes U(1)_{em}\cr}
}
The important point is that unlike the \sm, electroweak symmetry breaking
takes place with a Higgs that carries color.  Within the context of this model
the problem of fermion masses becomes the question of why the other five
quarks and leptons have small
nonzero masses (in relation to the electroweak symmetry breaking
scale).  We do not seek a fundamental solution to this problem
in this paper.  Instead we will add the usual \sm\ Higgs $\phi$ and imagine
that its VEV is fifty times smaller than the VEV of $H$.  This offers a partial
explanation of light fermion masses, and maybe renormalization explains why
$\vev{H}$ is much greater than $\vev{\phi}$ since the parameters for $H$ run
faster.
We write the \sm\ Yukawa couplings to generate these masses,
\eqn\oldyuk
{
L_0 = \lambda_3 \overline f_L \phi e_R + \lambda_4 \overline Q_L \phi d_R
       + \lambda_5 \overline Q_L \phi^c u_R
	+ \lambda' \overline T_L \phi b_R +{\rm h.c.} \ .
}
At this point we mention that all the hypercharge assignments have been
normalized in such a way that the \sm\ Higgs hypercharge is one,
i.e.~$\phi \sim (1, 2, 1)$ under eq.\gaugegrp.

In order to give both components of the top the correct electric
charge, we must have $y_{t_R}=2-2y_Q$ for $t_R$ in the $\overline{10}$ and
$y_{t_R}=(3-y_Q)/2$ for $t_R$ in the 10.
The electromagnetic charge operator is
\eqn\emop{
Q_{em}=I_3+\half(Y'+(1-3y_Q) T)
}
One can show that $Y = Y' + (1-3y_Q)T$ gives the familiar values for the
hypercharges of the color triplet quarks.  $T$ is an \suc 5\ generator in
$\tilde{U}(1)$ with
the following representation when acting on the five colors of ordinary
plus exotic quarks
\eqn\t
{
T = {\rm Diag}
	\({1 \over 3}, {1 \over 3}, {1 \over 3}, {-1 \over 2}, {-1 \over
								2}\)
\ .}

Further characteristics of the $\overline{10}$ and the 10 model differ
substantially and we will discuss each one separately in the following two
sections.

\newsec{ The $\overline{10}$ model} In this section we study in more detail the
$t_R\sim(\overline{10},1,2-2y_Q)$ model. When the $\fred\times\suc3$ singlet
piece of $\chi$ and $H$ gets a VEV we are left with $\fred\times\suc3\times
U(1)_{em}$ as the unbroken gauge group.  $\fred$ is the same exotic force
sector discussed in \fh\ and in much more detail in \chsb. The \suc 3\ singlet
quarks that come from the 5 of \suc5\ have charges  $(5y_Q-1\pm 2)/4$. These
quarks are confined by the \fred\ force and will lead to charge $0,\pm1,
10y_Q\pm\half, 10y_Q-{3\over2}$ exotic mesons.

The field $t_R$ in the $\overline{10}$ will contain a charge 2/3 color triplet
which is just the usual right handed top quark.  The exotics that come from the
$t_R$ have  charges \hbox{$(3-5y_Q)/2$} for the singlets and $(13-15y_Q)/12$
for the \fred\ doublet color
\hyphenation{anti-trip-lets} antitriplets, which will be bound into a charge-0
meson.

The $[\sul ]^2U(1)'$ anomaly equation implies $y_Q=1/5$.  Thus all the exotic
mesons will have integer charge. However since $t_R$ is in the $\overline{10}$
of \suc 5, cancellation of the  color anomaly is no longer automatic.  This
forces us to add exotics.  Recall that the anomaly of the 5 and 10 of \su 5\
are equal.  Our philosophy is to associate the higher dimensional
representations of \suc 5\ with the higher energy physics.  Thus we cancel the
anomaly by adding two electroweak singlets in the 10.  We also need to give
large mass to the exotic components of $t_R$ and the other exotic fields.
One possible exotic particle content is
\eqn\exotics{\eqalign{
a_R\sim(10,1,y_a) & \quad c_R\sim(10,1,y_c) \cr
E_R\sim (24,1,-2) \quad & P_R\sim(75,1,-y_a+2/5) \quad Q_R\sim(75,1,-y_c+2/5),
\cr}}
with the following Yukawa terms
\eqn\exyuk{ \eqalign{
&\lambda_t \chi \overline{E^c_R}t_R + \lambda_a \chi^c \overline{P^c_R}a_R +
\lambda_c \chi \overline{Q^c_R}c_R +
\cr
&\lambda_{\ss E} \rho_{\ss E} \overline{E^c_R}E_R +
 \lambda_{\ss P}\rho_{\ss P} \overline{P^c_R}P_R +
	\lambda_{\ss Q}\rho_{\ss Q} \overline{Q^c_R}Q_R + {\rm h.c.} \ .\cr
}}
It is straightforward to check that $a_R, c_R$ and the exotic components of
$t_R$ get masses of order the
\suc 5\ scale without affecting the mass of the ordinary color triplet quarks.
The $\rho$ Higgses are \suc5,\sul\ singlets whose $U(1)'$ charges are chosen so
as to allow the Yukawa coupling above.  We imagine these $\rho$'s getting a VEV
at the same time as $H$ thus giving $E_R,P_R,Q_R$ masses of order the top quark
mass.

We still have to cancel the $[\suc 5]^2U(1)'$ and $[U(1)']^3$ anomalies.
This will determine the hypercharges $y_a$ and $y_c$.  The equations that
correspond to cancelling the above two anomalies are
\eqn\anocan{\eqalign{
& {-118 \over 5} + 47(y_a+y_c) =  0 \cr
& {3752 \over 25} + 36(y_a+y_c)-90(y_a^2+y_c^2)+65(y_a^3+y_c^3) = 0 . \cr
}}
These equations can be combined to give a quadratic equation whose solutions
are irrational and are given approximately by $y_a=1.64$ and $y_c=-1.14$.

\newsec{ The 10 model}
We now consider $t_R$ in the 10 of \suc 5.  One key difference between the ten
and ten bar model is that $H$ also breaks the \fred\ subgroup of \suc5.  Thus
the unbroken gauge group is the same as in the standard model.  Since the
coupling in \fred\ is given by the strong coupling constant, the three heavy
gauge bosons from that sector will be slightly heavier that the $Z$.  Also
these \fred\ gauge bosons will not couple to ordinary matter at tree level so
that their production in hadron colliders will be suppressed.

This model has two right handed top quarks with charge 2/3 that couple to the
broken \fred\ generators.
The singlet and anti-triplet exotics of $t_R$ have charges $(5y_Q+1)/4$ and
$(13-15y_Q)/12$, respectively.  As in the ten bar model, the $[\sul ]^2U(1)'$
anomaly equation implies $y_Q=1/5$. While exotics are no longer needed to
cancel the color anomaly, we need them to cancel the $U(1)'$ anomalies and
to give the exotic components of $t_R$
a mass.  Thus we add the following particle content
\eqn\exten{
E_R\sim(75,1,-1) \quad P_R\sim(24,1,y_P) \quad Q_R\sim(75,1,y_Q)
}
with the following Yukawa terms
\eqn\yukten{
\lambda_{t} \chi^c \overline{E^c_R}t_R +
\lambda_{\ss E} \rho_{\ss E} \overline{E^c_R}E_R +
\lambda_{\ss P}\rho_{\ss P} \overline{P^c_R}P_R +
\lambda_{\ss Q}\rho_{\ss Q} \overline{Q^c_R}Q_R
+ {\rm h.c.}  \ .
}
The $\rho$ Higgses behave in the same way as they did in the ten bar model.

The values for $y_P$ and $y_Q$ are determined from anomaly cancellation, which
can be summarized in the following equations:
\eqn\anot{
\eqalign{
& 47-10y_P-50y_Q =  0 \cr
& {1067 \over 25} -24y_P^3-75y_Q^3 =  0 .\cr}
}
This has one real irrational solution given approximately by $y_P=-1.67$ and
$y_Q=1.27$.

\newsec{Conclusions}  We have presented two models which lead to a heavy
top quark within the framework of extended color models \fh. 

In our approach, the first two generations are exact copies of each
other.  However the third-generation right-handed top quark is in a
nonfundamental representation of the extended color group.  It is this property
which we use to characterize the physics at the higher energy scale.

Both models lead to new particles at the color symmetry breaking scale.  Even
without a detailed calculation, one expects that this scale will satisfy the
same experimental bounds of the original \suc5\ model \fh.  These bounds can be
as low as 300 GeV \gs--\oh.  Certain theoretical prejudices may put it at the
1-10 TeV scale \chsb, yet the exotic \fred\ force sector has been sufficiently
changed in our ``ten'' model, that the constraints from \chsb\ do not apply.
In any case we find it more interesting to consider the possibility of a
multi-GeV color breaking scale, which would then naturally lead to our exotics
having masses below a TeV. The ten model in particular will have a  top-like
right handed quark with the same electric charge, which gets its mass by
coupling to the exotic sector only.

However, independant of how high one chooses the color symmetry breaking scale,
both models predict that the $E_R,P_R,Q_R$ exotics will have masses of order
the top quark mass.  There are obvious ways around this prediction, such as
choosing a different mass generating mechanism than the one suggested here, or
considering a different exotic particle content.

Another feature of our models is the irrational electric charges of some of the
exotic particles.  This was forced upon us by anomaly cancellation, and by the
desire to keep the number of new fields to a minimum.  Irrational
charges certainly run against current folklore, though it leads to no
contradiction with accelerator experiments.  We do not know if irrational
charges
are a generic feature of such models or if perhaps some cleverer model
builder will be able to construct an exotic sector with rational charges only.

Unfortunately our model offers no insights into the origin of mass for the
lighter generations, nor does it explain  why higher dimensional
representations should involve higher energy physics.  Instead we present this
model as an existence proof for \sm-like theories in which the third generation
is treated differently than the first two.  The models we presented are
consistent with experiment and lead to new physics at the 100 GeV to 1 TeV
scale.  We hope these models can serve as a possible direction of study into
the origin of fermion masses.

\centerline{\bf Acknowledgements} I would especially like to thank Jim Cline,
for many useful discussions and for helping me trace down my algebra mistakes.
Special thanks are also due Cliff Burgess and Robert Foot. I would like to
acknowledge fruitful conversations with Jean-Ren\'e Cudell,
Keith Dienes, C.~S.~Lam, Bernie
Margolis, and the rest of the McGill High Energy Theory group. Finally  I would
also like to thank Gordon Kane whose talk at the CINVESTAV Workshop on High
Energy Phenomenology in Mexico City, 1991, started me thinking about
this problem.
\listrefs\bye